\documentclass[aps,prd,singlecolumn,nofootinbib,superscriptaddress]{revtex4-2}

\usepackage{amsmath,amssymb,graphicx,physics,bm}
\usepackage{hyperref}
\usepackage{graphicx}
\usepackage{subcaption}

\begin{document}

\title{Controlled Quantum Metrology with Anisotropic Heisenberg Spin Interactions under Intrinsic Decoherence}

\author{S.K. Singh}
\affiliation{Department of Physics, Akal University, Talwandi Sabo, Bathind-151302, India}
\email{singhshailendra3@gmail.com}
\author{Jia-Xin Peng}
\affiliation{School of Physical Science and Technology, Nantong University, Nantong 226019, People’s Republic of China}
\author{Y-J Zhu}
\affiliation{Institute of Quantum Science and Technology, Yanbian University, Yanji,Jilin, China}
\author{Mohammad Khalid}
\affiliation{James Watt School of Engineering, University of Glasgow,  G128QQ, UK}
\affiliation{University Centre for Research and Development, Chandigarh University, Mohali, Punjab, 140413, India}

\date{\today}

\begin{abstract}
We theoretically investigate quantum parameter estimation in a two-qubit anisotropic Heisenberg spin system with Dzyaloshinskii-Moriya (DM) interaction in the presence of intrinsic decoherence described by the Milburn model. Using the Quantum Fisher Information (QFI), we study the estimation of both the uniform magnetic field and the DM interaction strength. Analytical expressions for the time-evolved density matrix are obtained and used to explore the effects of exchange anisotropy, intrinsic decoherence, and probe-state preparation on the achievable estimation precision. Our results show that suitable tuning of the anisotropic exchange coupling and the initial entangled state can considerably enhance the estimation performance, with different optimal parameter regimes emerging for magnetic-field and DM-interaction sensing. To better understand the role of quantum resources in metrology, we also examine the behaviour of concurrence, quantum coherence, and von Neumann entropy. Overall, our findings demonstrate that anisotropic Heisenberg spin systems with DM interaction provide a promising and flexible platform for high-precision quantum metrology even in the presence of intrinsic decoherence.

\end{abstract}

\maketitle

\section{Introduction}

Quantum metrology has emerged as one of the most promising applications of quantum information science, enabling parameter estimation with precisions that surpass classical limits through the exploitation of quantum coherence, entanglement, squeezing, and other nonclassical resources \cite{1,2,3,4,5}. Owing to these advantages, quantum-enhanced sensing has found widespread applications in atomic clocks, magnetometry, interferometry, gravitational-wave detection, and precision measurements. The ultimate achievable precision is fundamentally bounded by the quantum Cram\'er--Rao inequality, in which the QFI quantifies the maximum information that can be extracted about an unknown parameter encoded in a quantum state \cite{6}. In realistic implementations, however, unavoidable decoherence and environmental noise degrade quantum coherence and consequently reduce the attainable estimation precision \cite{7,8}. To overcome these limitations, considerable effort has been devoted to developing robust metrological protocols based on monitored quantum trajectories, dynamical decoupling, feedback control, and critical quantum sensing \cite{9,10,11,12,13}.

Among the various physical platforms employed for quantum technologies, Heisenberg spin systems constitute an attractive framework because of their rich quantum properties and their direct relevance to condensed-matter physics and spin-based information processing. In particular, anisotropic exchange interactions together with the DM interaction arising from spin-orbit coupling provide additional degrees of freedom for manipulating coherence and quantum correlations. Recent investigations have demonstrated that DM interactions can significantly improve parameter estimation and quantum information processing under external magnetic fields \cite{14,15}. Motivated by these developments, we investigate quantum metrology in an anisotropic two-qubit Heisenberg spin system under Milburn intrinsic decoherence. Using the Quantum Fisher Information as the principal figure of merit, we analyze the estimation of both the uniform magnetic field and the DM interaction strength while examining the influence of exchange anisotropy and probe-state preparation. Our results show that suitable engineering of the system parameters and the initial entangled probe state can substantially enhance metrological performance and provide useful insights for the realization of robust spin-based quantum sensing protocols under intrinsic decoherence.

Motivated by these considerations, in this work we study quantum parameter estimation in an anisotropic XYZ Heisenberg spin system with DM interaction in the presence of intrinsic decoherence. Our main focus is to understand how well the uniform magnetic field and the DM interaction strength can be estimated using the Quantum Fisher Information. We also explore how the exchange anisotropy, intrinsic decoherence, and the choice of the initial entangled state affect the estimation precision. As we will show, careful tuning of these parameters can lead to a significant improvement in the sensing performance, with different optimal conditions emerging for the estimation of different physical quantities. These findings provide a better understanding of spin-based quantum metrology and may be useful for the development of robust quantum sensing schemes in realistic noisy environments. 

The manuscript is organized as follows. In Section II, we give the Model Hamiltonian for anisotropic  Heisenberg spin model with DM interaction and briefly describe the intrinsic decoherence model used in this work. Section III presents the theoretical framework of QFI and single parameter Gaussian quantum metrology. In Section IV, we present and discuss the numerical results, including the effects of exchange anisotropy, magnetic field, DM interaction, probe-state preparation, intrinsic decoherence, and the role of various quantum resources in relation to the metrological performance. To better understand the role of quantum resources in metrology, we also examine the quantum correlations and von Neumann entropy in this Section. Finally, we summarised our findings in Section V.

\section{Model Hamiltonian and Intrinsic Decoherence}

We consider a two-qubit anisotropic  Heisenberg spin model in the presence of a DM interaction directed along the $z$ axis and an uniform magnetic-field parameter $b$ acting along the z-direction. The Hamiltonian of this system is given as \cite{16}

\begin{equation}
H
=
J_x\,\sigma_x^{(1)}\sigma_x^{(2)}
+
J_y\,\sigma_y^{(1)}\sigma_y^{(2)}
+
J_z\,\sigma_z^{(1)}\sigma_z^{(2)}
+
D_z
\left(
\sigma_x^{(1)}\sigma_y^{(2)}
-
\sigma_y^{(1)}\sigma_x^{(2)}
\right)
+
b
\left(
\sigma_z^{(1)}
+
\sigma_z^{(2)}
\right),
\label{Ham}
\end{equation}

where $J_x$, $J_y$, and $J_z$ denote the exchange coupling strengths along the three spatial directions, $D_z$ represents the strength of the Dzyaloshinskii--Moriya interaction arising from spin--orbit coupling, and $b$ denotes the external magnetic-field parameter.

In the computational basis
$\{|00\rangle,|01\rangle,|10\rangle,|11\rangle\}$,
the Hamiltonian assumes the matrix form

\begin{equation}
H=
\begin{pmatrix}
J_z+2b & 0 & 0 & J_x-J_y \\
0 & -J_z & J_x+J_y+2iD_z & 0 \\
0 & J_x+J_y-2iD_z & -J_z & 0 \\
J_x-J_y & 0 & 0 & J_z-2b
\end{pmatrix}.
\label{HamMatrix}
\end{equation}

The Hamiltonian possesses a block-diagonal structure, separating naturally into the subspaces
$\{|00\rangle,|11\rangle\}$
and
$\{|01\rangle,|10\rangle\}$.
Consequently, its eigenvalues can be obtained analytically.

For the $\{|00\rangle,|11\rangle\}$ sector, the eigenenergies are

\begin{equation}
E_{1,2}
=
J_z
\pm
\Omega_1,
\end{equation}

where

\begin{equation}
\Omega_1
=
\sqrt{
4b^2
+
(J_x-J_y)^2
}.
\end{equation}

Similarly, for the $\{|01\rangle,|10\rangle\}$ sector, one obtains

\begin{equation}
E_{3,4}
=
-
J_z
\pm
\Omega_2,
\end{equation}

with

\begin{equation}
\Omega_2
=
\sqrt{
(J_x+J_y)^2
+
4D_z^2
}.
\end{equation}

Hence, the energy spectrum is governed by two characteristic frequencies: $\Omega_1$, determined by the combined influence of the magnetic field and exchange anisotropy, and $\Omega_2$, controlled by the interplay between the exchange interaction and the Dzyaloshinskii--Moriya coupling.

To incorporate intrinsic decoherence, we adopt the Milburn model, whose master equation is

\begin{equation}
\frac{d\rho(t)}{dt}
=
-i[H,\rho(t)]
-
\frac{\gamma}{2}
[H,[H,\rho(t)]],
\label{Milburn}
\end{equation}

where $\gamma$ is the intrinsic decoherence rate. The first term generates coherent unitary dynamics, while the second term describes the gradual suppression of coherence through stochastic phase diffusion.

Expressing the initial density matrix in the eigenbasis of the Hamiltonian, the formal solution of Eq.~(\ref{Milburn}) can be written as

\begin{equation}
\rho(t)
=
\sum_{m,n}
e^{-i(E_m-E_n)t}
e^{-\frac{\gamma t}{2}(E_m-E_n)^2}
\rho_{mn}(0)
|m\rangle
\langle n|,
\label{rhoformal}
\end{equation}

where
$\rho_{mn}(0)=\langle m|\rho(0)|n\rangle$ and the Gaussian damping factor explicitly shows that coherences between energy eigenstates decay more rapidly when their energy separation is larger.

We consider a parameterized two-qubit probe state that continuously interpolates between Bell-type maximally entangled states and a separable superposition state. The degree of entanglement is controlled by the parameter $\theta$, with maximal entanglement occurring at $\theta=0$ and $\theta=\pi/2$, while the state becomes separable at $\theta=\pi/4$. This parameterization enables us to investigate about the choice of the initial probe state influences on the achievable metrological precision.

\begin{equation}
|\psi(0)\rangle
=
\frac{\cos\theta}{\sqrt{2}}
\left(
|00\rangle+|11\rangle
\right)
+
\frac{\sin\theta}{\sqrt{2}}
\left(
|01\rangle+|10\rangle
\right),
\label{ProbeState}
\end{equation}

where the mixing angle $\theta$ determines the relative contribution of the two Bell-type components. The corresponding initial density operator is

\begin{equation}
\rho(0)
=
|\psi(0)\rangle
\langle\psi(0)|.
\end{equation}

The time-evolved density matrix obtained from Eq.~(\ref{rhoformal}) serves as the starting point for evaluating the Quantum Fisher Information associated with the estimation of the magnetic-field parameter $b$ and the DM interaction strength $D_z$. This analytical framework enables us to systematically investigate how exchange anisotropy, probe-state preparation, and intrinsic decoherence jointly influence the attainable metrological precision.

\section{Single Parameter Gaussian Metrology}

The central objective of quantum metrology is to estimate an unknown physical parameter with the highest possible precision by exploiting quantum resources. In the present work, the metrological performance of the anisotropic Heisenberg spin system is quantified through the Quantum Fisher Information (QFI), which provides the ultimate bound on the achievable estimation accuracy.

According to the quantum Cram\'er--Rao inequality, the variance associated with an unbiased estimator of a parameter $\lambda$ satisfies \cite{16,17}

\begin{equation}
\Delta \lambda
\ge
\frac{1}
{\sqrt{\nu F_Q(\lambda)}},
\label{QCRB}
\end{equation}

where $\nu$ denotes the number of independent measurements and $F_Q(\lambda)$ is the Quantum Fisher Information corresponding to the parameter $\lambda$. Throughout this work, we have consider two estimation tasks, namely the determination of the uniform magnetic-field parameter $b$ and DM interaction strength $D_z$.

The intrinsic decoherence dynamics discussed in the previous section transforms the initial probe state into a mixed state described by the density operator $\rho(\lambda)$, where the dependence on the parameter $\lambda$ is encoded through the Hamiltonian evolution. Let

\begin{equation}
\rho(\lambda)
=
\sum_{i}
p_i
|\psi_i\rangle
\langle\psi_i|
\end{equation}

be the spectral decomposition of the density matrix, with $\lambda_i$ and $|\psi_i\rangle$ representing its eigenvalues and eigenvectors, respectively. QFI for a mixed state is then given by

\begin{equation}
F_Q(\lambda)
=
2
\sum_{i,j}
\frac{
\left|
\langle
\psi_i
|
\partial_\lambda
\rho
|
\psi_j
\rangle
\right|^2
}
{p_i+p_j},
\label{QFI}
\end{equation}

where the summation extends over all pairs satisfying
$\lambda_i+\lambda_j>0$.
This expression provides a general measure of the sensitivity of the quantum state to infinitesimal variations of the estimated parameter.

In the present investigation, the derivative
$\partial_\lambda \rho$
is evaluated numerically by means of a central finite-difference approximation which is given as
$\partial_\lambda \rho\approx\frac{\rho(\lambda+\delta\lambda)
-\rho(\lambda-\delta\lambda)}{2\delta\lambda}$,
where the finite-difference step is chosen as $\delta b = 10^{-6}$ for the estimation of the uniform magnetic-field parameter and $\delta D_z = 10^{-4}$ for the estimation of the DM interaction strength. In the evaluation of Eq.~(\ref{QFI}), only terms satisfying $\lambda_i+\lambda_j>10^{-12}$ are included in the summation to avoid numerical instabilities associated with nearly vanishing denominators. For the calculations of the maximum QFI, the optimization is carried out over the interval $0\le t\le5$ using a uniform grid of 300 time points. This numerical procedure provides a common computational framework for estimating both the magnetic-field parameter $b$ and the DM interaction strength $D_z$.

We would also like to point here that unlike conventional analyses based on a fixed probe state, here we have taken the parametrized two-qubit probe state that interpolates between Bell type entangled states and a separable superposition state as  introduced in Eq.~(\ref{ProbeState}) which also allow the probe preparation to be optimized through the mixing angle $\theta$. Consequently, the achievable metrological precision depends not only on the intrinsic decoherence rate and the Hamiltonian parameters $(J_x,J_y,J_z,b,D_z)$ but also on the structure of the initial quantum state.

The QFI obtained from Eq.~(\ref{QFI}) is employed throughout the remainder of this work to investigate the effects of exchange anisotropy, probe-state engineering, intrinsic decoherence, and the DM interaction on quantum parameter estimation. Particular attention is given to identifying the parameter regimes in which the QFI is maximized, thereby revealing the optimal conditions for high-precision estimation of the uniform magnetic field and the DM interaction strength.

\subsection{Quantum Resources}

To better understand the metrological behaviour of the system, we also analyze the concurrence, the $l_{1}$-norm quantum coherence, and the von Neumann entropy of the evolved state.

The concurrence is evaluated as
\[
C(\rho)=\max\left\{0,\sqrt{\lambda_1}-\sqrt{\lambda_2}-\sqrt{\lambda_3}-\sqrt{\lambda_4}\right\},
\]
where $\lambda_i$ are the eigenvalues, in decreasing order, of the matrix
\[
\rho(\sigma_y\otimes\sigma_y)\rho^{*}(\sigma_y\otimes\sigma_y).
\]

The $l_{1}$-norm quantum coherence is defined as
\[
C_{l_1}(\rho)=\sum_{i\neq j}|\rho_{ij}|,
\]
which quantifies the total magnitude of the off-diagonal elements of the density matrix in the computational basis.

The von Neumann entropy is given by
\[
S(\rho)=-\mathrm{Tr}(\rho\log_2\rho)
       =-\sum_i p_i\log_2 p_i,
\]
where $p_i$ are the eigenvalues of the density matrix. It provides a measure of the mixedness of the quantum state.

\section{Results and Discussion}

In this section, we investigate the metrological performance of the proposed anisotropic Heisenberg spin model under intrinsic decoherence through the QFI associated with the estimation of the uniform magnetic field $b$ and the DM interaction strength $D_z$. We have given particular emphasis to investigate the effects of exchange anisotropy $J_{y}$, probe-state preparation and intrinsic decoherence which ultimately determine the achievable estimation precision which will lead to enhanced quantum sensing.

\begin{figure}[t]
\centering
\subfloat[]{\includegraphics[width=0.48\textwidth]{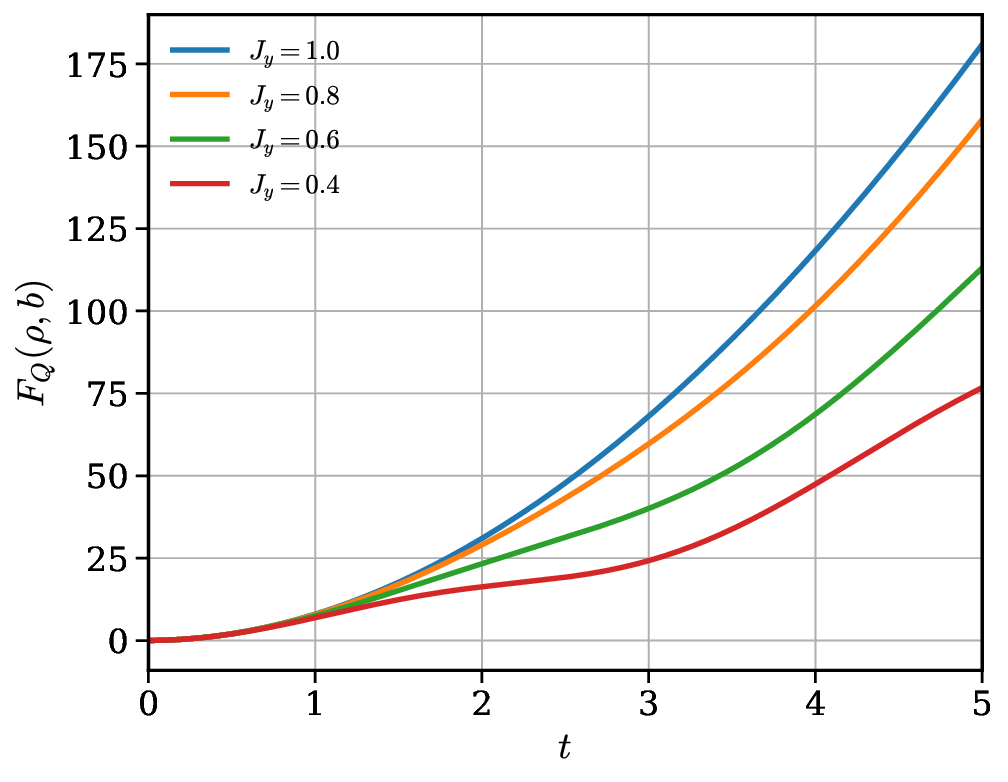}
\label{fig:1a}}
\hfill
\subfloat[]{\includegraphics[width=0.48\textwidth]{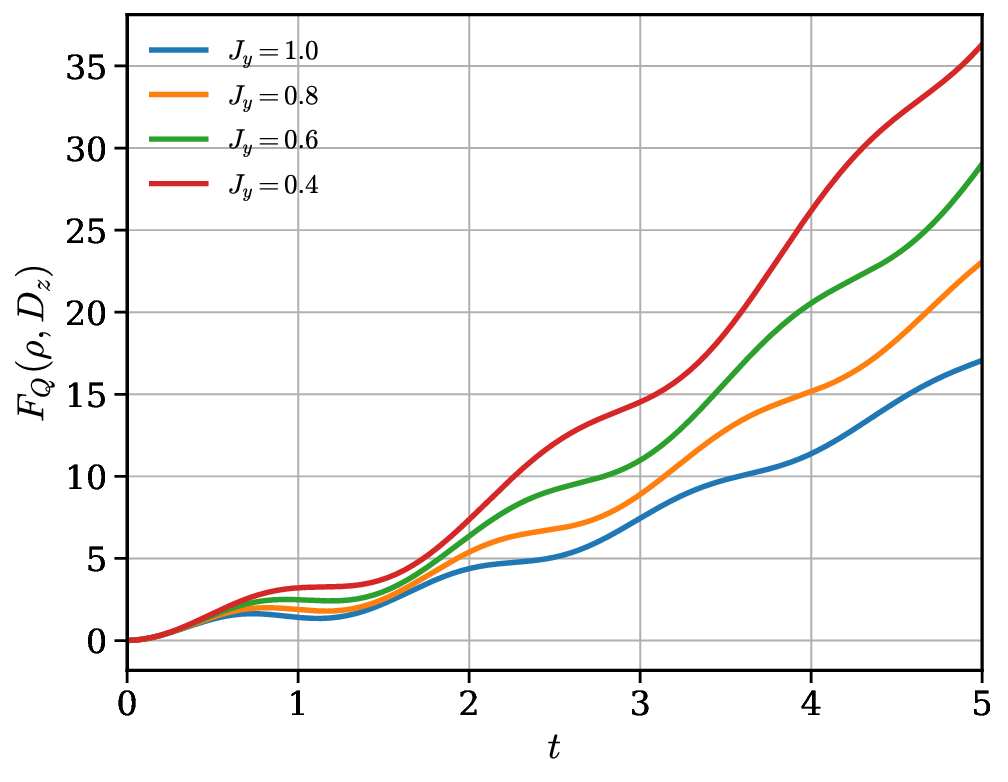}
\label{fig:1b}}
\caption{Time evolution of $F_Q(\rho,b)$ and $F_Q(\rho,D_z)$ for the estimation of (a) the uniform magnetic field $b$ and (b) the DM interaction strength $D_z$, respectively, for different values of $J_y$. The remaining parameters are fixed at $J_x=1.0$, $J_z=0.2$, $D_z=0.5$, $b=0.3$, $\theta=\pi/4$, and $\gamma=0.02$.}
\label{fig:1}
\end{figure}

The time evolution of $F_Q(\rho,b)$ and $F_Q(\rho,D_z)$ is shown in Fig.~\ref{fig:1a} and Fig.~\ref{fig:1b}, respectively, for different values of the anisotropic exchange coupling $J_y$. It can be seen that $F_Q(\rho,b)$ increases steadily with time for all values of $J_y$ as shown in Fig.~\ref{fig:1a} which means the sensitivity of the probe state to the uniform magnetic field gradually improves as the dynamics evolve. In addition, the magnitude of $F_Q(\rho,b)$ increases significantly with a gradual increase in $J_y$, where $J_y=1.0$ consistently provides the highest values throughout the evolution. This suggests that stronger exchange coupling along the $y$-direction enhances the precision of magnetic-field estimation and helps preserve the metrological advantage even in the presence of intrinsic decoherence. The behaviour of $F_Q(\rho,D_z)$ plotted in Fig.~\ref{fig:1b} is qualitatively different and opposite in nature. Although it also grows with time, its dependence on the anisotropy parameter is reversed. In particular, gradually reducing $J_y$ leads to a significant enhancement in $F_Q(\rho,D_z)$, with the largest values obtained for $J_y=0.4$, whereas the isotropic case $J_y=1.0$ yields the lowest estimation precision for the DM interaction. Therefore, the exchange anisotropy plays distinct roles depending on the parameter being estimated: larger values of $J_y$ are beneficial for improving $F_Q(\rho,b)$, whereas smaller values are more favourable for enhancing the precision of $F_Q(\rho,D_z)$. This contrasting behaviour highlights the important role of the anisotropic exchange interaction $J_y$ as a practical control parameter for optimizing the metrological performance of the system depending on the desired estimation protocol.\

\begin{figure}[t] 
\centering 
\subfloat[]{\includegraphics[width=0.48\textwidth]{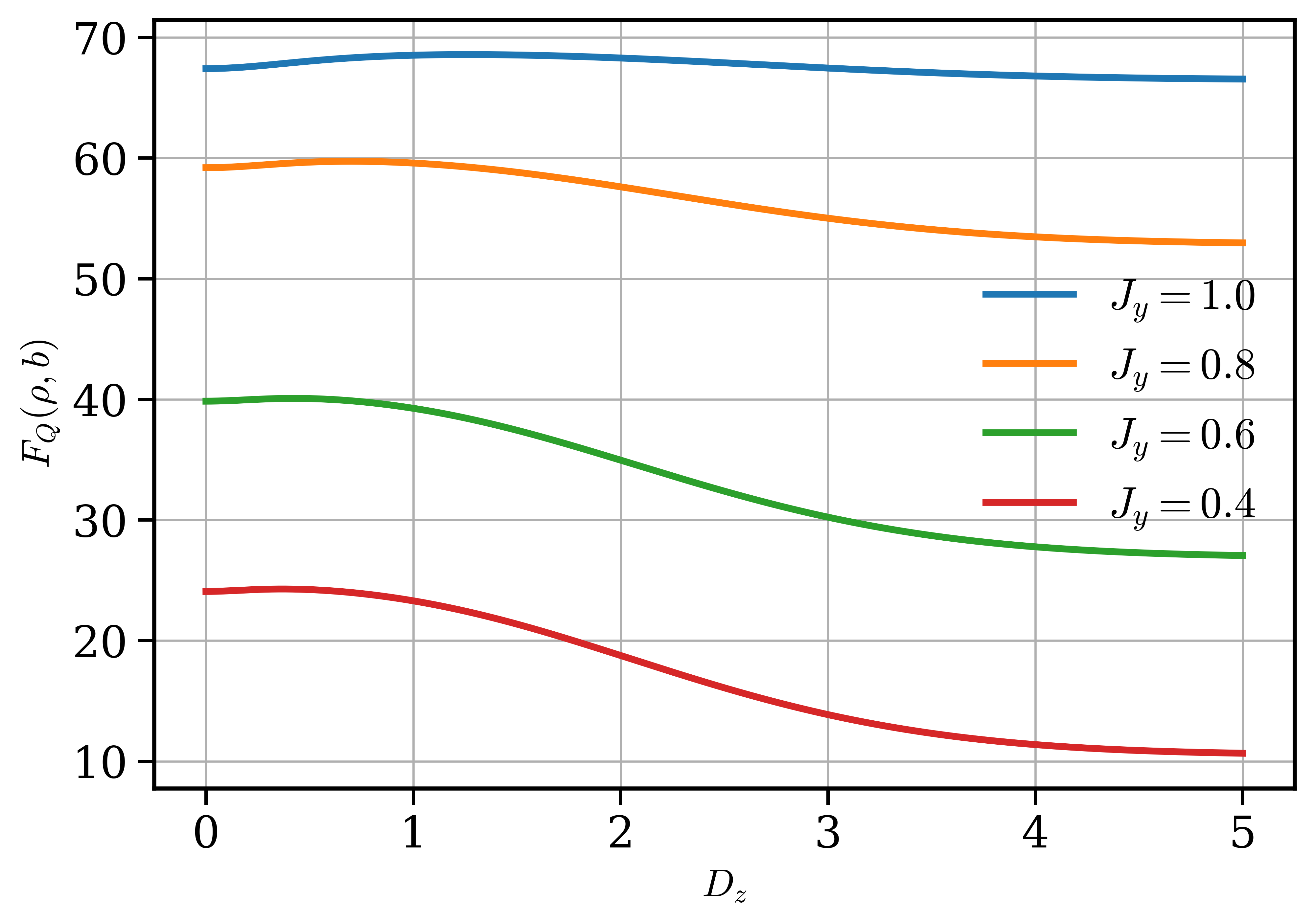} 
\label{fig:2a}} 
\hfill 
\subfloat[]{\includegraphics[width=0.48\textwidth]{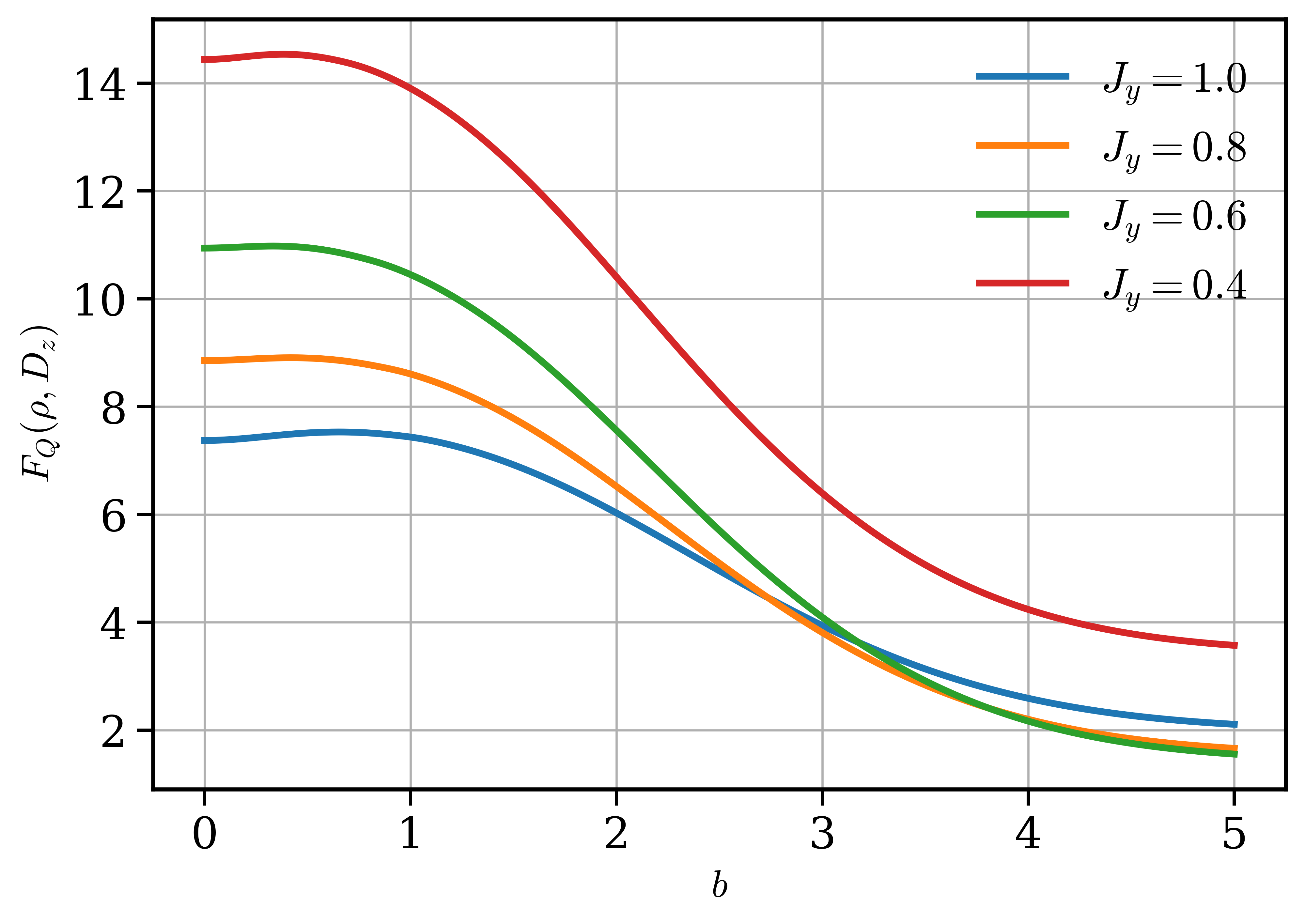} 
\label{fig:2b}} 
\caption{Dependence of (a) $F_Q(\rho,b)$ on the DM interaction strength $D_z$ and (b) $F_Q(\rho,D_z)$ on the uniform magnetic field $b$ for different values of $J_y$. All other parameters are the same as used in Fig.~1.} 
\label{fig:2} 
\end{figure}

To further investigate the quantum metrological properties of the proposed spin model, we now examine how the estimation precision is affected by the DM interaction strength and the uniform magnetic field, as shown in Fig.~\ref{fig:2}. For all considered values of the exchange anisotropy parameter $J_y$, it can be seen from Fig.~\ref{fig:2a} that $F_Q(\rho,b)$ decreases gradually with increasing $D_z$, indicating that a stronger DM interaction weakens the precision of magnetic-field estimation. However, the rate of decrease depends strongly on $J_y$. For $J_y=1.0$, $F_Q(\rho,b)$ remains relatively robust and exhibits only a small reduction as $D_z$ increases, whereas for $J_y=0.4$ the decline is much more pronounced, giving rise to the steepest slope among all the curves. Therefore, larger values of $J_y$ not only provide higher values of $F_Q(\rho,b)$ but also make the estimation process less sensitive to variations in the DM interaction strength. This behaviour highlights the important role of exchange anisotropy in controlling and stabilizing the metrological performance of the system. A completely different trend is observed in Fig.~\ref{fig:2b}, where $F_Q(\rho,D_z)$ is plotted as a function of the uniform magnetic field $b$. In this case, the highest estimation precision is obtained for $J_y=0.4$, with $F_Q(\rho,D_z)$ attaining its largest values in the low-field region. However, as the magnetic field strength increases, $F_Q(\rho,D_z)$ decreases steadily for all values of $J_y$, and the reduction is particularly pronounced for $J_y=0.4$. As a result, the gap between the different curves gradually narrows, leading to crossover behaviour at larger values of $b$. These results indicate that although smaller values of $J_y$ are advantageous for estimating the DM interaction at low magnetic fields, this advantage gradually diminishes as the field strength increases.

Overall, these results show a nontrivial interplay among the uniform magnetic field $b$, the DM interaction strength $D_z$, and the exchange anisotropy parameter $J_y$ to determine the metrological performance of the system Hamiltonian. This demonstrate that careful tuning of these parameters can selectively enhance the estimation precision of the desired quantity, making the proposed anisotropic Heisenberg spin model a flexible platform for controlled quantum metrology.

\begin{figure}[t]
\centering
\subfloat[]{\includegraphics[width=0.48\textwidth]{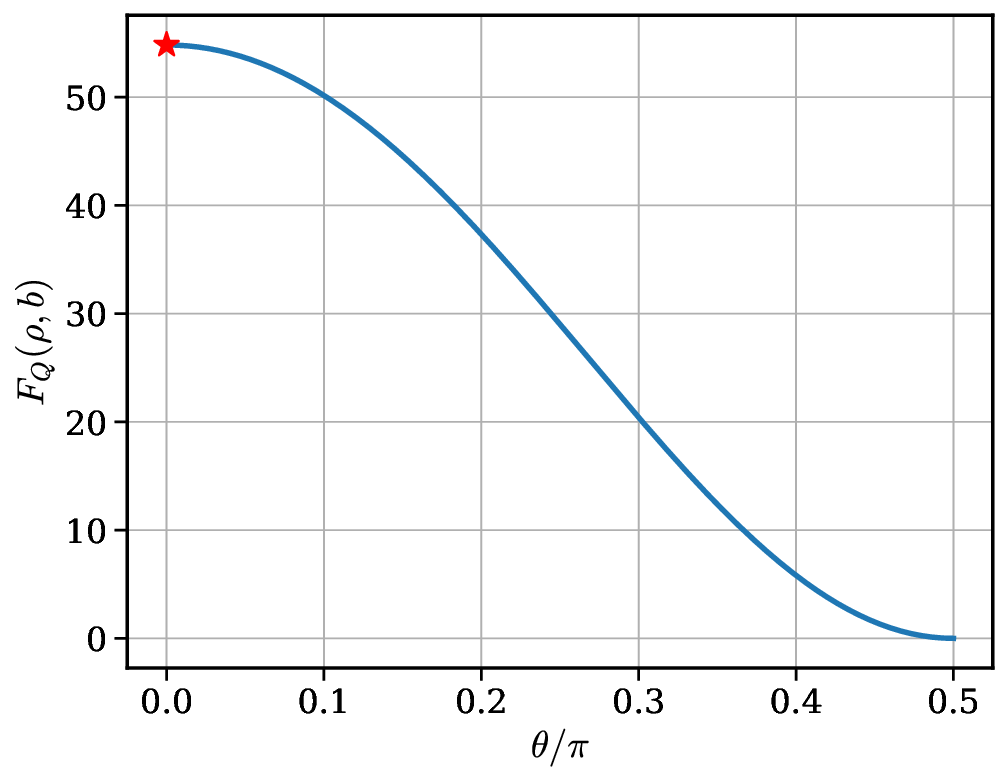}
\label{fig:3a}}
\hfill
\subfloat[]{\includegraphics[width=0.48\textwidth]{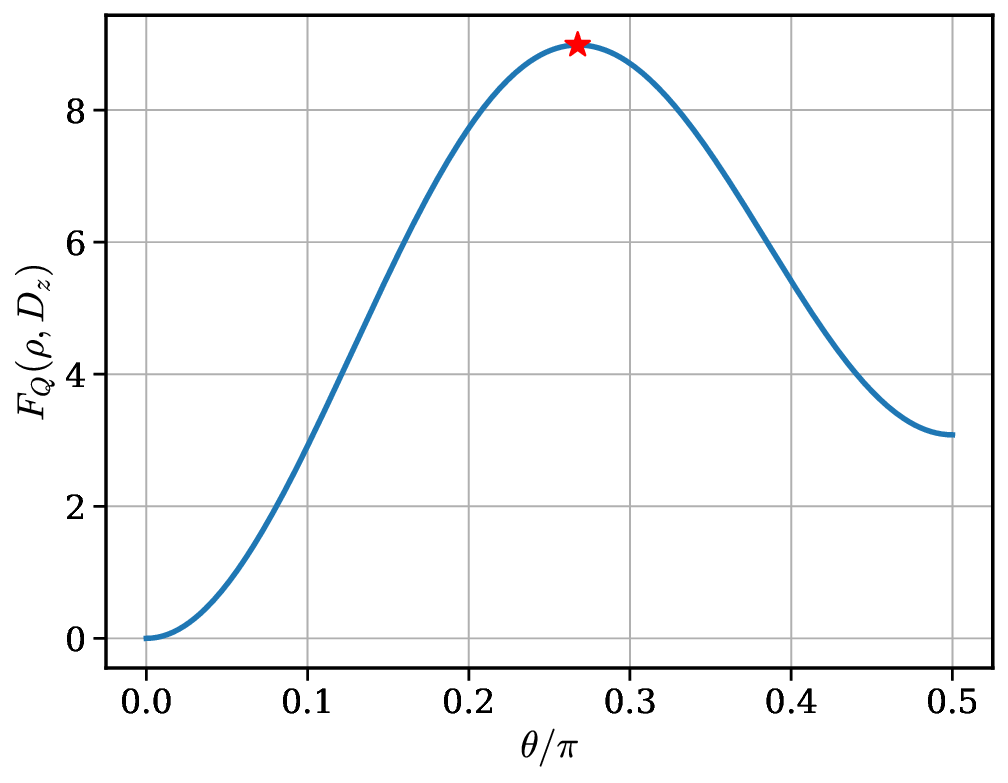}
\label{fig:3b}}
\caption{Dependence of (a) $F_Q(\rho,b)$ and (b) $F_Q(\rho,D_z)$ on the initial-state parameter $\theta$. All other parameters are the same as those used in Fig.~1, with $t=3.0$.}
\label{fig:fig3}
\end{figure}

In addition to exchange anisotropy coupling $J_y$, probe-state engineering is another significant method for enhancing quantum metrology in this  system.  We  analyse the dependence of QFI with the parameter $\theta$ in Fig.~\ref{fig:fig3}, which demonstrate that the choice of the initial state plays a crucial role in determining the achievable estimation precision.
We first consider the estimation of the uniform magnetic field $b$, as shown in Fig.~\ref{fig:3a}. It is evident that $F_Q(\rho,b)$ attains its maximum value at $\theta=0$ which corresponds to the Bell state employed in the present work, and decreases steadily afterwards as $\theta$ increases and ultimately become nearly to zero as $\theta$ approaches $\pi/2$. This result clearly indicates that the Bell state serves as the most effective probe for estimating the magnetic field $b$ whereas on gradually moving away from this initial state reduces the achievable estimation precision.
Fig.~\ref{fig:3b} shows the dependence of $F_Q(\rho,D_z)$ on the probe-state parameter $\theta$. Unlike the behaviour observed for $F_Q(\rho,b)$, the value of $F_Q(\rho,D_z)$ starts from zero at $\theta=0$ with $\theta$ and gradually reaches a clear maximum around $\theta/\pi \approx 0.27$ then gradually decreases for larger values of $\theta$. This means that the Bell state is not the most suitable choice for estimating the DM interaction strength and a partially entangled initial state provides the highest estimation precision. These results indicate that the optimal probe state depends on the parameter being estimated and emphasize the importance of selecting an appropriate initial state to achieve the best metrological performance in this model.

\begin{figure}[t]
\centering
\subfloat[]{\includegraphics[width=0.48\textwidth]{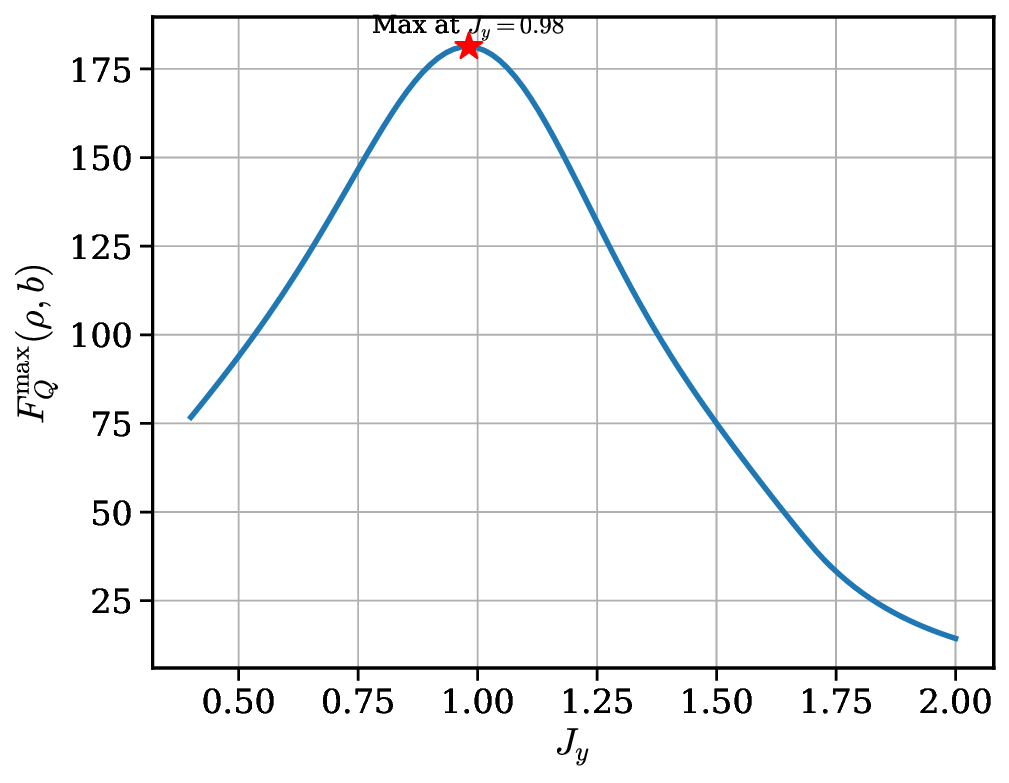}
\label{fig:4a}}
\hfill
\subfloat[]{\includegraphics[width=0.48\textwidth]{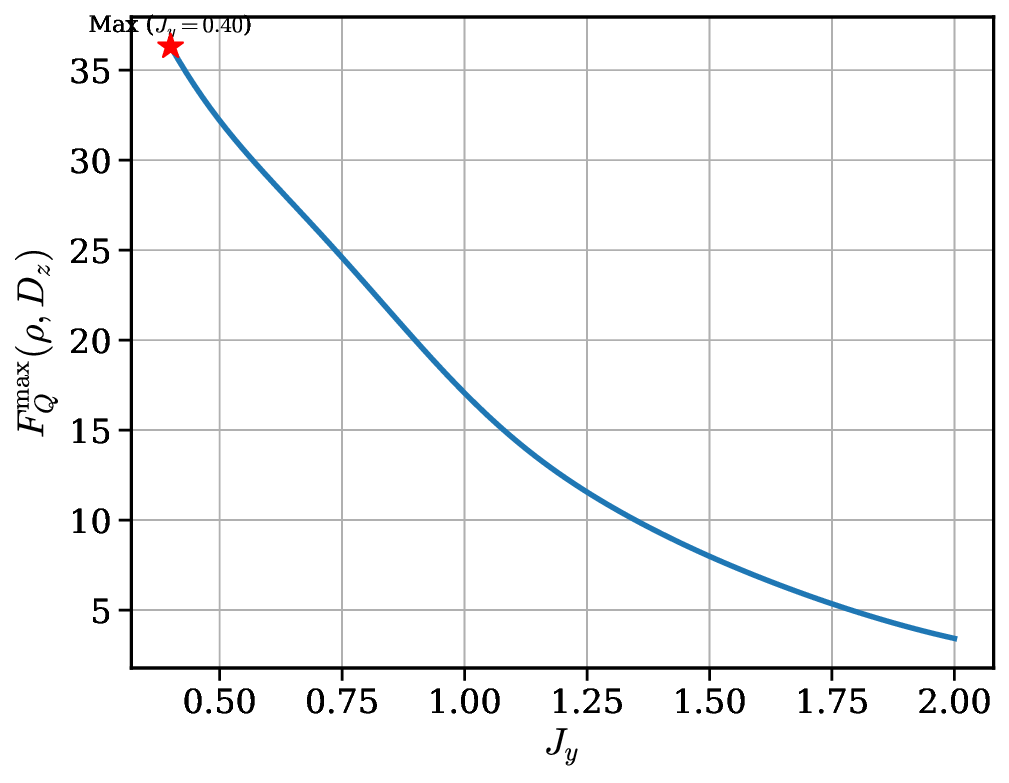}
\label{fig:4b}}
\caption{Maximum values of (a) $F_Q^{\max}(\rho,b)$ and (b) $F_Q^{\max}(\rho,D_z)$ as functions of the exchange coupling parameter $J_y$. All other parameters are same like used in Fig.~1, with the maximum taken over the time interval $0 \leq t \leq 5$.}
\label{fig:fig4}
\end{figure}

To explore more in details about the role of exchange anisotropy, now we have plotted the maximum values of $F_Q(\rho,b)$ and $F_Q(\rho,D_z)$ with the exchange coupling parameter $J_y$, as shown in Fig.~\ref{fig:fig4}, where as compared to full time evolution these plots highlight the best estimation precision that can be achieved for each value of $J_y$. It can be seen from Fig.~\ref{fig:4a} that $F_Q^{\max}(\rho,b)$ does not vary monotonically with $J_y$ and it increases steadily hence ultimately reaches its maximum value around $J_y \approx 0.98$ afterwards it decreases rapidly as $J_y$ is increased further. This indicates that the highest precision for magnetic field estimation is obtained at an intermediate value of the exchange anisotropy rather than at either extreme. The corresponding behaviour of $F_Q^{\max}(\rho,D_z)$ in Fig.~\ref{fig:4b} is much simpler as its largest value is obtained at $J_y \approx 0.40$, after which it decreases continuously with increasing $J_y$. In other words, smaller values of the exchange anisotropy are more favourable for estimating the DM interaction strength whereas increasing $J_y$ gradually reduces the achievable precision.

\begin{figure}[t]
\centering
\subfloat[]{\includegraphics[width=0.48\textwidth]{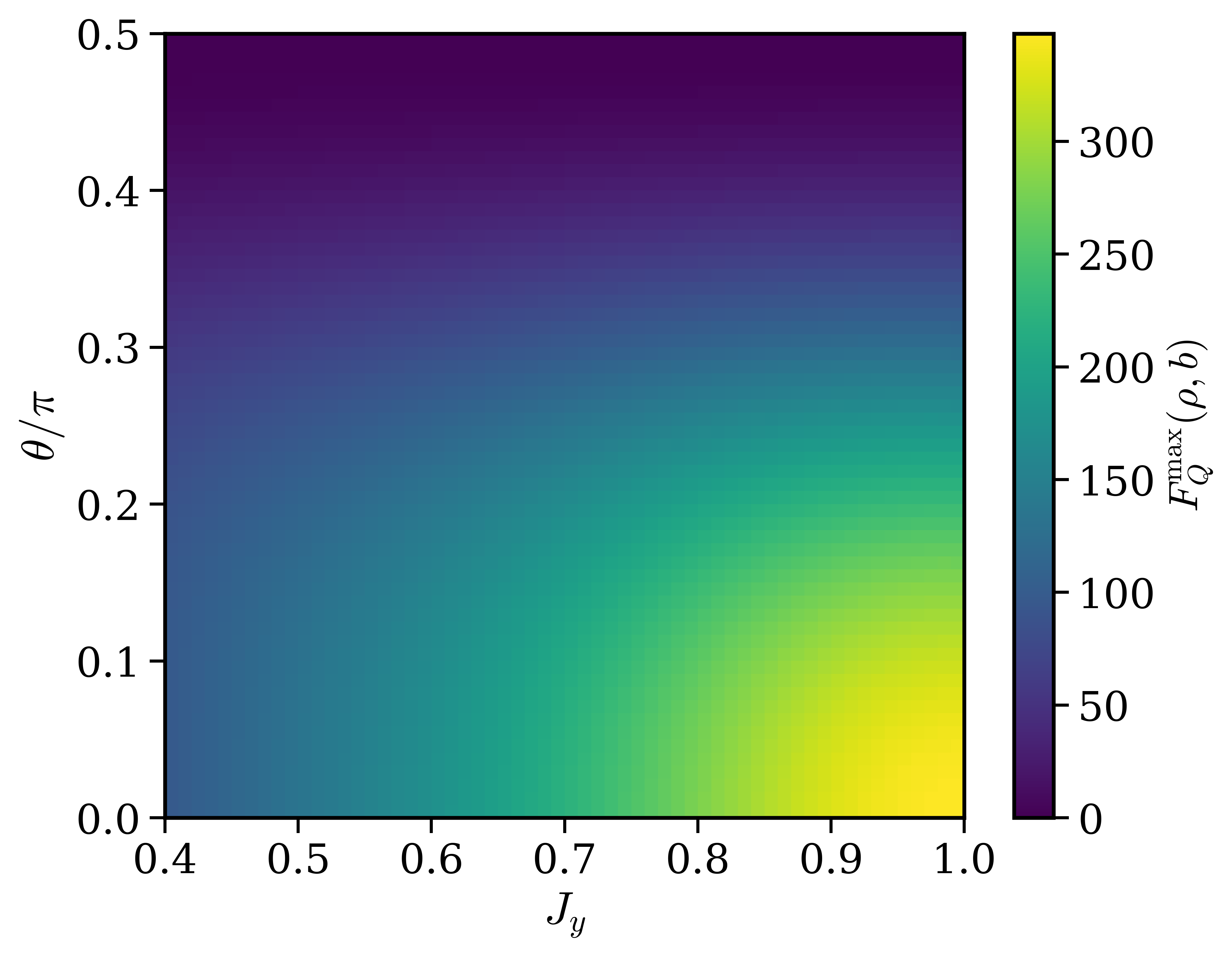}
\label{fig:5a}}
\hfill
\subfloat[]{\includegraphics[width=0.48\textwidth]{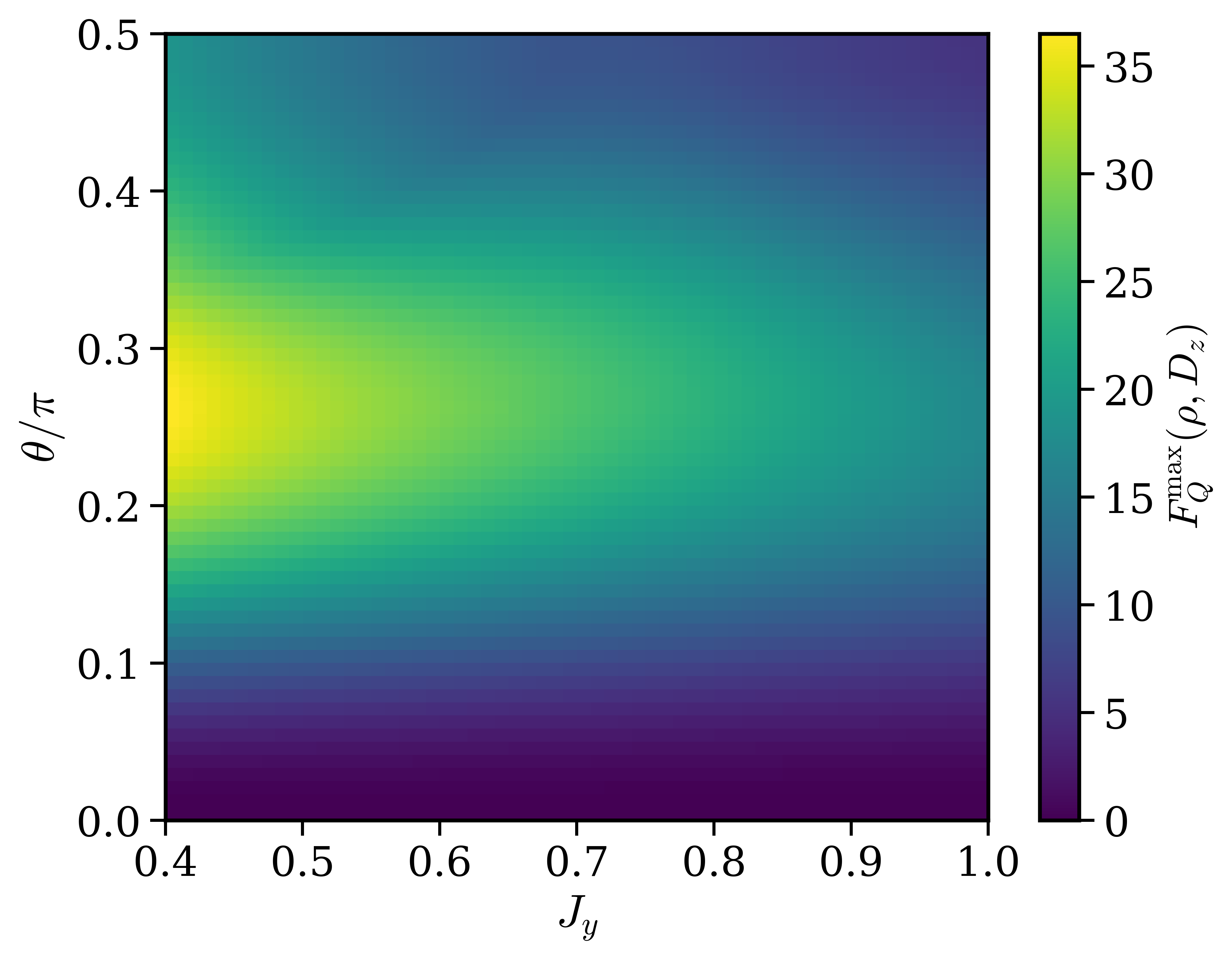}
\label{fig:5b}}
\caption{Two-dimensional density plots of (a) $F_Q^{\max}(\rho,b)$ and (b) $F_Q^{\max}(\rho,D_z)$ in the $(J_y,\theta)$ parameter plane. The remaining parameters are the same as those used in Fig.~1, with the maximum taken over the time interval $0 \leq t \leq 5$.}
\label{fig:fig5}
\end{figure}

Now we examine the combined effect of the exchange anisotropy parameter $J_y$ and the probe-state parameter $\theta$ on the maximum achievable values of $F_Q(\rho,b)$ and $F_Q(\rho,D_z)$ through the density plots shown in Fig.~\ref{fig:fig5}. We can see from Fig.~\ref{fig:5a} that the highest values of $F_Q^{\max}(\rho,b)$ are obtained when $\theta$ is close to zero and $J_y$ is close to 1.0. on gradually moving away from either of these conditions leads to a gradual reduction in the achievable precision which indicates that magnetic-field estimation benefits from a Bell-state preparation together with relatively large values of $J_y$ whereas Fig.~\ref{fig:5b} shows a very different pattern for $F_Q^{\max}(\rho,D_z)$. In this case, the highest values are found for smaller values of $J_y$ which is around 0.4 and for a probe-state parameter near $\theta/\pi \approx 0.27$. As either $J_y$ or $\theta$ moves away from this region, $F_Q^{\max}(\rho,D_z)$ decreases steadily. Therefore, the optimal conditions for estimating the DM interaction are quite different from those required for magnetic-field estimation. Overall, these results highlight the importance of jointly tuning $J_y$ and $\theta$ to achieve the best metrological performance for the parameter of interest.

\begin{figure}[t]
\centering
\subfloat[]{\includegraphics[width=0.48\textwidth]{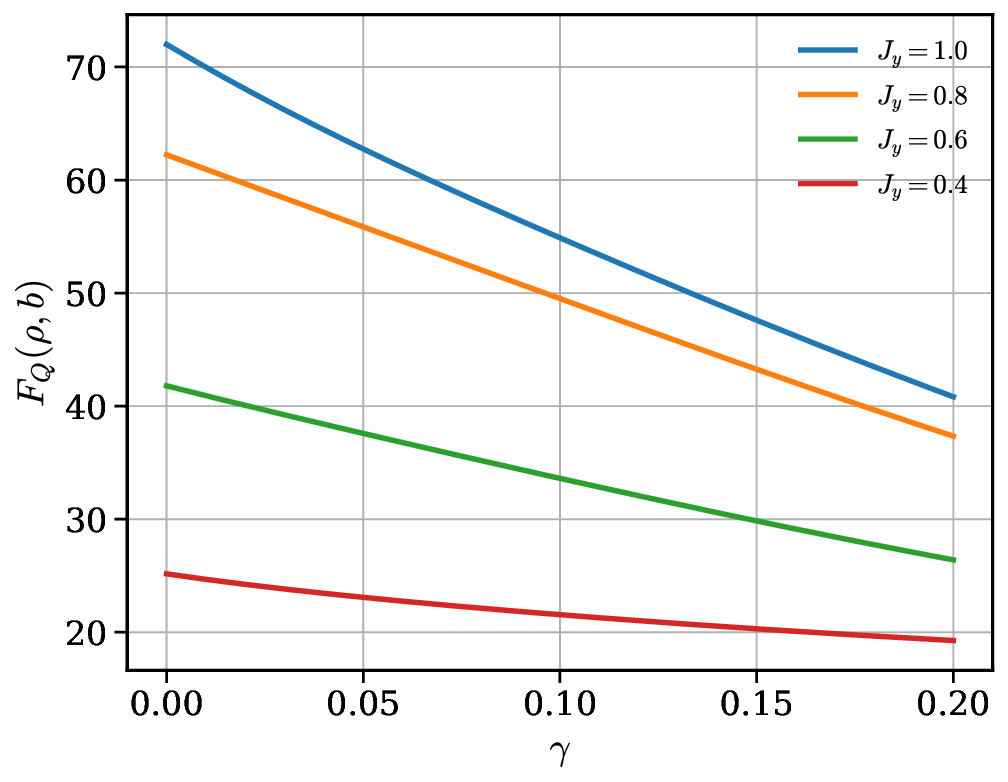}
\label{fig:6a}}
\hfill
\subfloat[]{\includegraphics[width=0.48\textwidth]{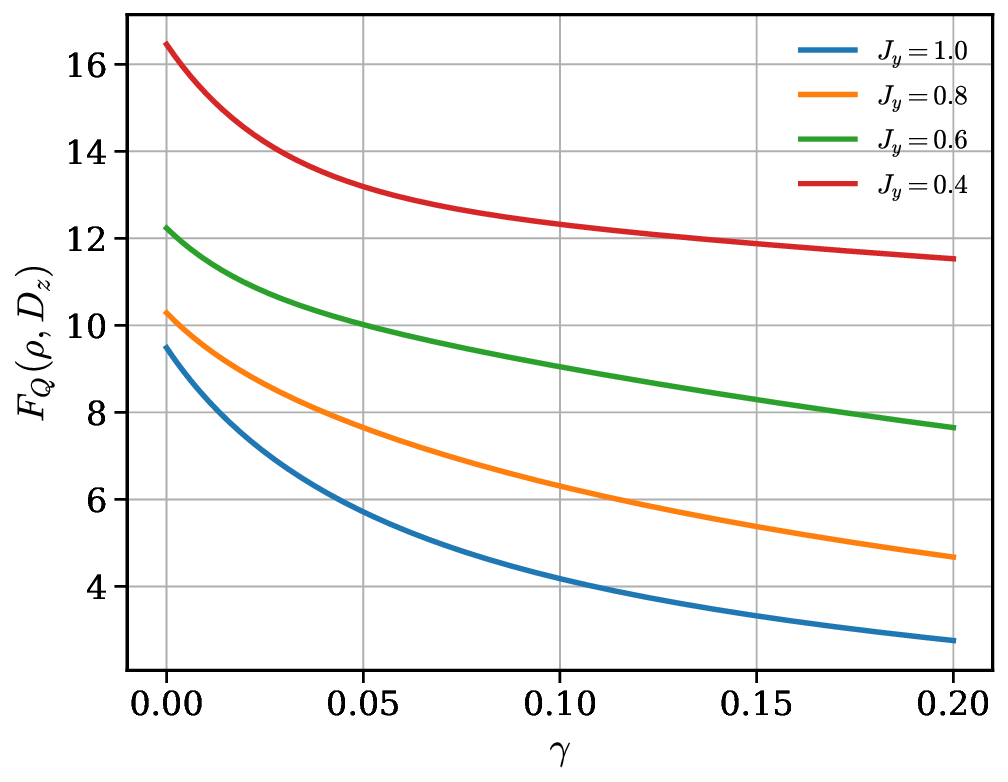}
\label{fig:6b}}
\caption{Dependence of (a) $F_Q(\rho,b)$ and (b) $F_Q(\rho,D_z)$ on the intrinsic decoherence rate $\gamma$ for different values of $J_y$. All other parameters are the same as those used in Fig.~1, with $t=3.0$.}
\label{fig:fig6}
\end{figure}

Now we investigate the influence of intrinsic decoherence on the estimation precision in Fig.~\ref{fig:fig6}, which illustrates the variation of QFI with the decoherence rate $\gamma$ for different values of the exchange coupling $J_y$. The variation of $F_Q(\rho,b)$ with the intrinsic decoherence rate $\gamma$ for different values of $J_y$ is plotted in Fig.~\ref{fig:6a} where it is observed that it decreases as the decoherence rate increases. This indicates that a intrinsic decoherence gradually degrades the estimation precision. The highest values are obtained for $J_y=1.0$, but this curve also exhibits the fastest decline with increasing $\gamma$. In comparison, the curves corresponding to smaller values of $J_y$ decrease more slowly, with $J_y=0.4$ showing the weakest dependence on decoherence, although its overall estimation precision remains lower. The corresponding behaviour of $F_Q(\rho,D_z)$ is given in Fig.~\ref{fig:6b}. We can see that $F_Q(\rho,D_z)$ also decreases monotonically as $\gamma$ increases for all values of $J_y$ which means that intrinsic decoherence adversely affects the estimation of the DM interaction as well. Unlike the magnetic-field case, all the curves experience a noticeable drop with increasing decoherence, the ordering remains unchanged throughout the considered range. Overall, these results show that intrinsic decoherence deteriorates the estimation precision for both the parameters although the choice of exchange anisotropy plays a crucial role in determining which sensing task can be optimized.

\begin{figure}[t]
\centering
\includegraphics[width=0.48\textwidth]{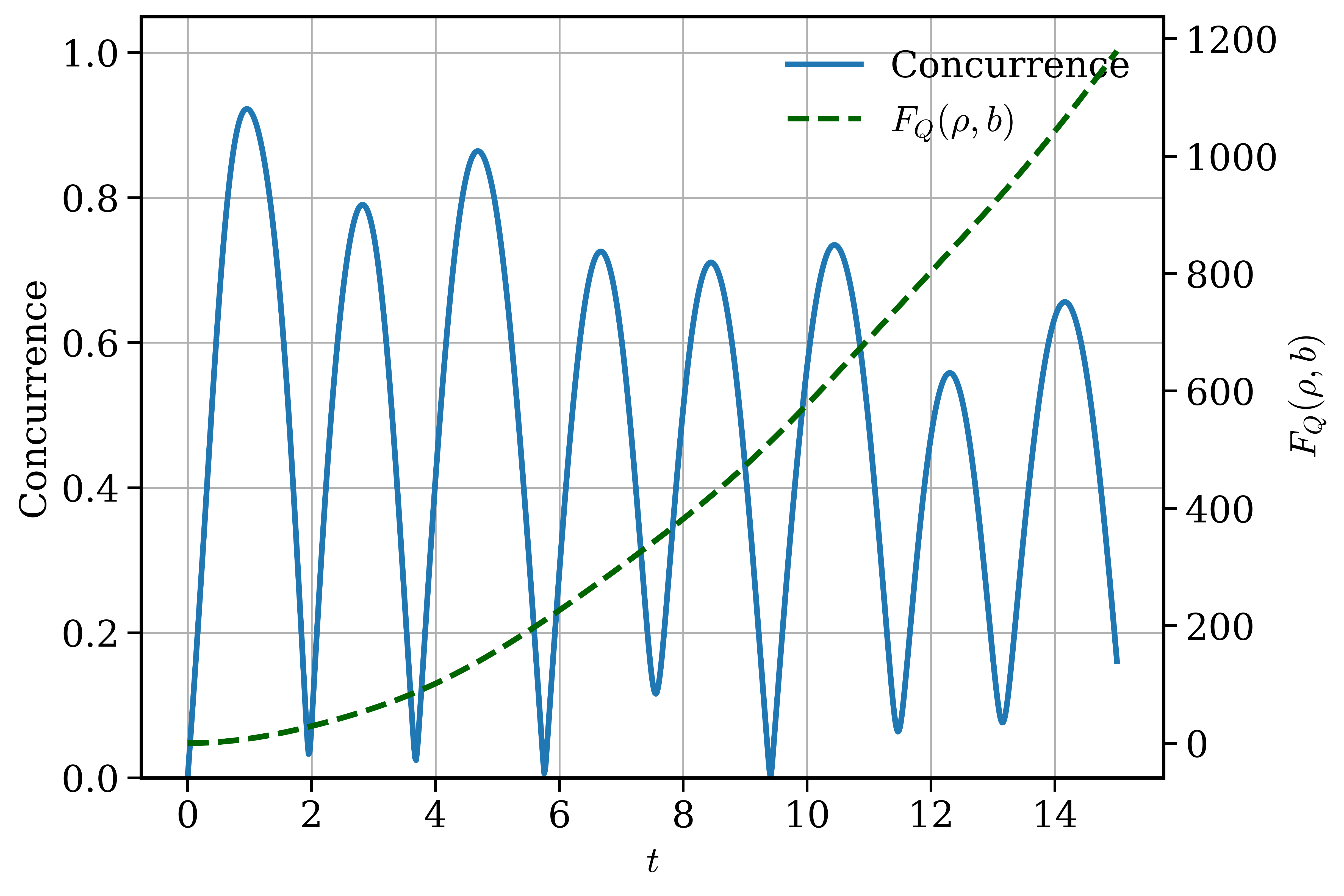}
\hfill
\includegraphics[width=0.48\textwidth]{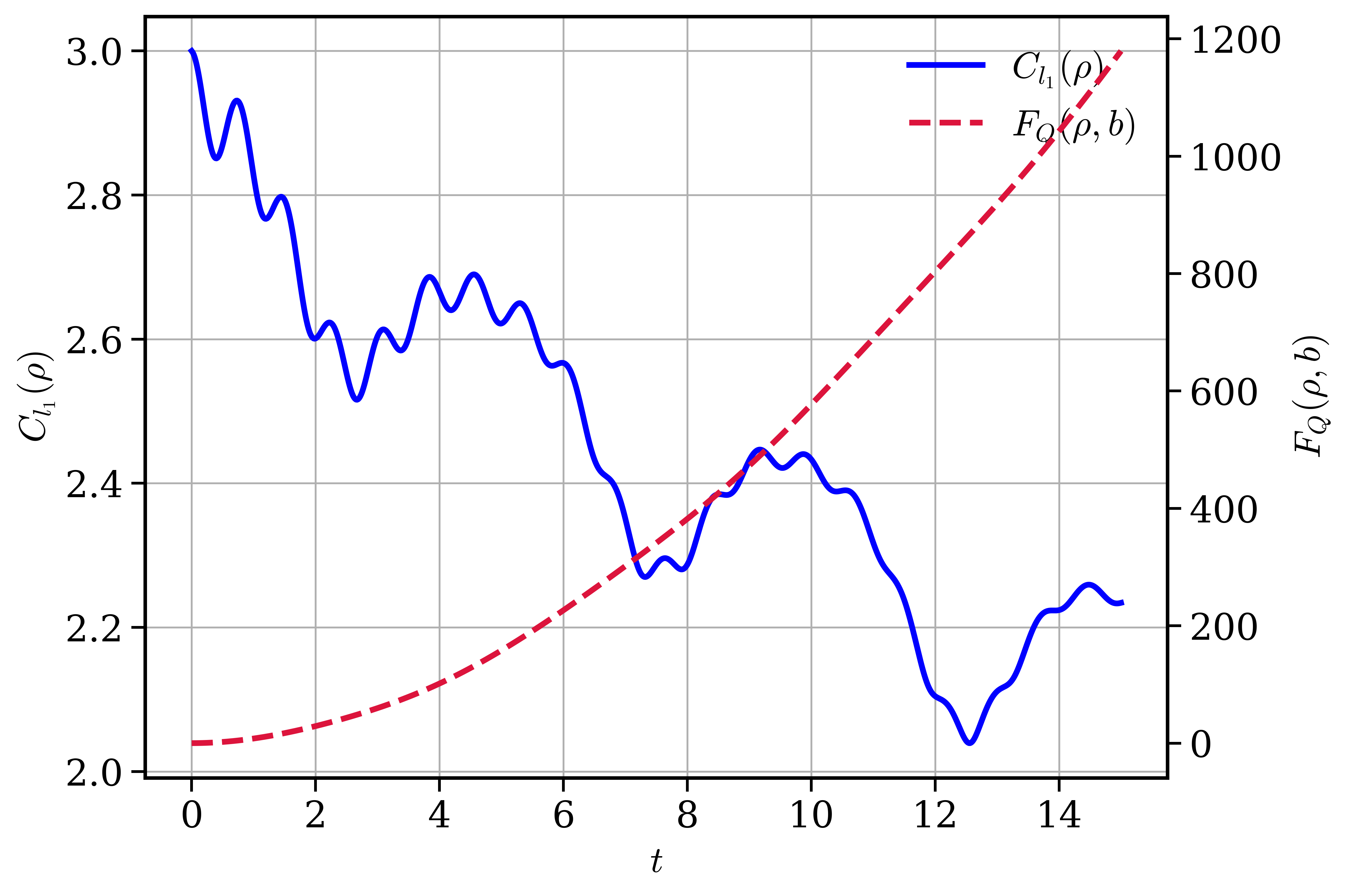}
\caption{Comparison of (a) concurrence and $F_Q(\rho,b)$, and (b) the $l_{1}$-norm quantum coherence and $F_Q(\rho,b)$ as functions of time. The parameter $J_y$ is fixed at $0.8$ whereas all other parameters are the same as those used in Fig.~1.}
\label{fig:Concurrence_Coherence_QFI}
\end{figure}

To better understand the role of quantum resources in metrology, we have compared the time evolution of the concurrence and the $l_{1}$-norm coherence with $F_Q(\rho,b)$, as shown in Fig.~\ref{fig:Concurrence_Coherence_QFI}. The concurrence exhibits repeated collapses and revivals phenomena which reflects the nontrivial entanglement dynamics of the system as plotted in Fig.~\ref{fig:Concurrence_Coherence_QFI}(a) whereas $F_Q(\rho,b)$ follows a very different trend and remains relatively large even when the concurrence becomes small. This clearly shows that higher estimation precision should  not be necessarily associated with strong bipartite entanglement and that concurrence alone cannot be used to predict the behavior of $F_Q(\rho,b)$. A similar comparison is presented in Fig.~\ref{fig:Concurrence_Coherence_QFI}(b), where the $l_{1}$-norm coherence is plotted together with $F_Q(\rho,b)$. The coherence gradually decreases with time, with only weak oscillations during the evolution. In contrast, $F_Q(\rho,b)$ does not follow this decay and continues to exhibit a different dynamical behavior. This indicates that a reduction in basis-dependent coherence does not automatically lead to a loss of estimation precision. Overall, these results suggest that neither concurrence nor the $l_{1}$-norm coherence has a simple one-to-one relationship with $F_Q(\rho,b)$. Instead, the achievable precision is determined by the overall response of the quantum state to changes in the parameter being estimated. The same qualitative behaviour is observed when the DM interaction strength $D_z$ is taken as the parameter of interest. In this case, $F_Q(\rho,D_z)$ also shows no direct one-to-one correspondence with either the concurrence or the $l_{1}$-norm coherence. Although these quantum resources evolve differently with time, the estimation precision is governed by the overall parameter sensitivity of the quantum state rather than by the amount of entanglement or coherence alone. This further confirms that neither concurrence nor basis-dependent coherence serves as a universal indicator of metrological performance in the present anisotropic Heisenberg spin system.

\begin{figure}[t]
\centering
\includegraphics[width=0.48\textwidth]{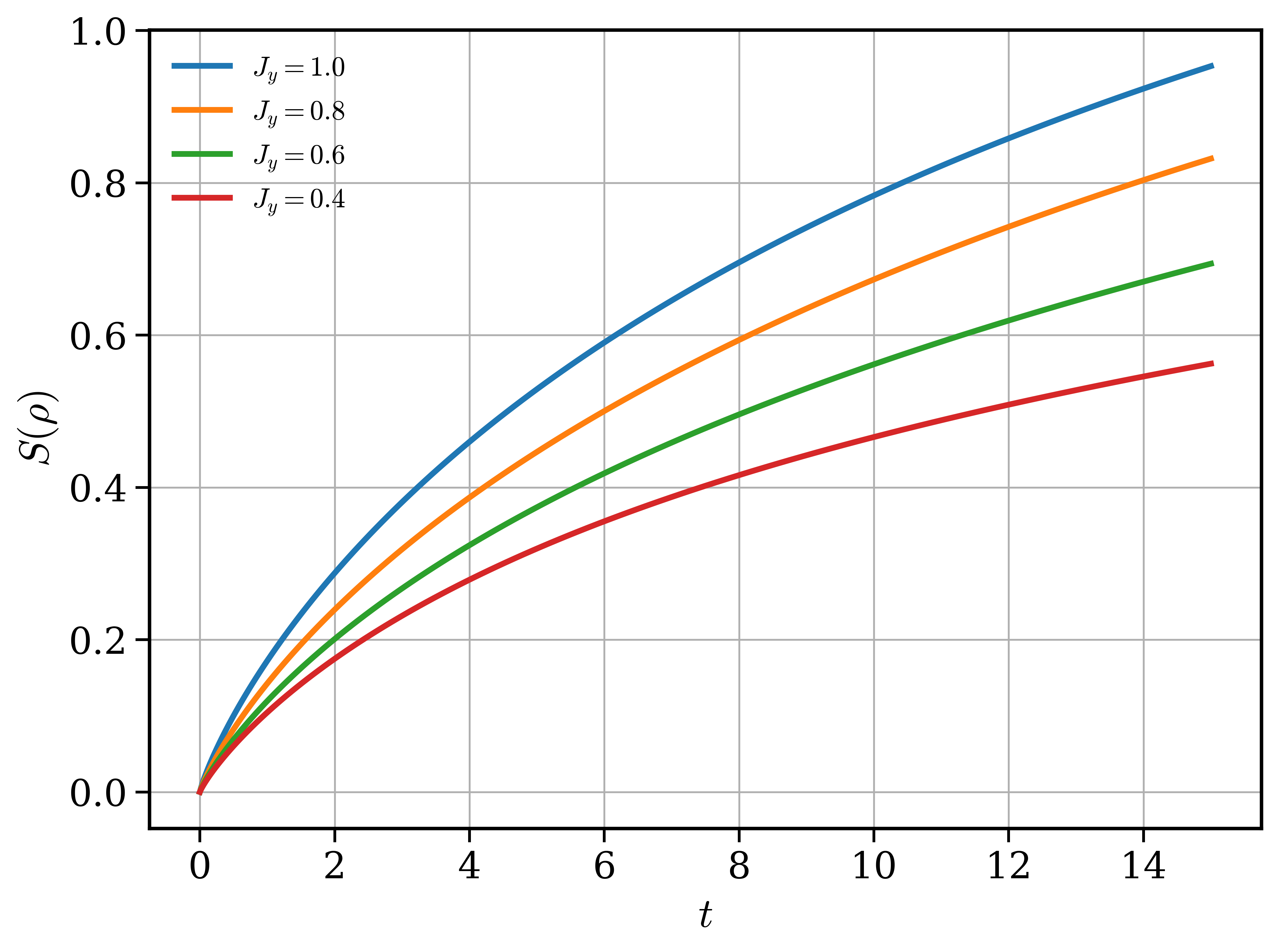}
\hfill
\includegraphics[width=0.48\textwidth]{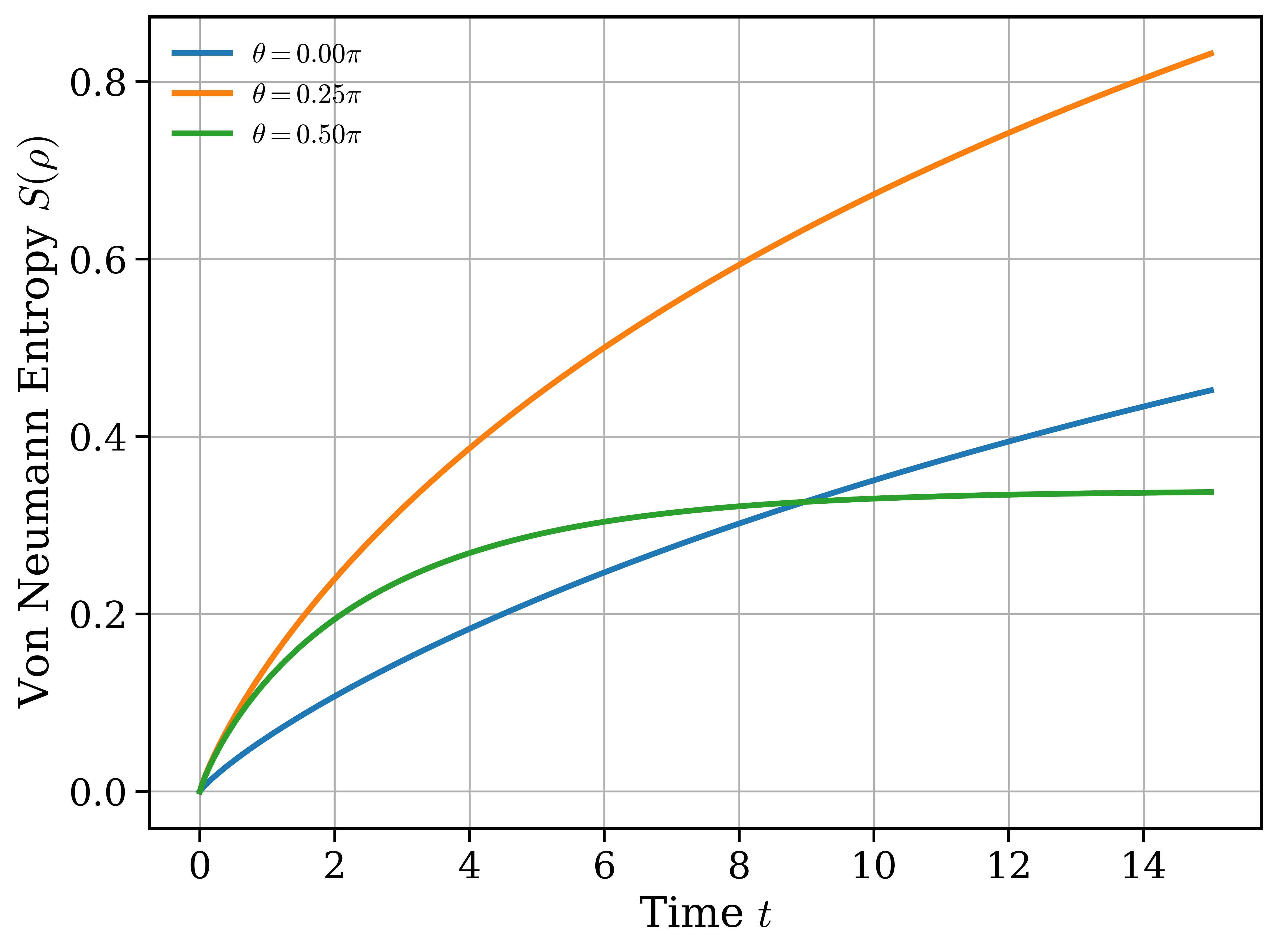}
\caption{Time evolution of the von Neumann entropy $S(\rho)$ for (a) different values of the exchange anisotropy parameter $J_y$ with $\theta=\pi/4$ and (b) different values of the probe-state parameter $\theta$ with $J_y=0.8$. All other parameters remain same like Fig.~1.}
\label{fig:Entropy_QFI}
\end{figure}

To further understand the behaviour of the Quantum Fisher Information, we also examine the von Neumann entropy of the evolving quantum state. The results shown in Fig.~\ref{fig:Entropy_QFI} help us explore how the mixedness generated by intrinsic decoherence is related to the metrological performance of the system. It can be seen from Fig.~\ref{fig:Entropy_QFI}(a) that the entropy increases gradually with time for all values of $J_y$, which means that the initial pure state becomes more mixed as the system evolves with time. We observe that larger values of $J_y$ produce higher entropy throughout the time evolution where $J_y=1.0$ gives the largest values and $J_y=0.4$ the smallest. We also notice that this behaviour is similar to that of $F_Q(\rho,b)$, where larger values of $J_y$ lead to better estimation precision. However, the trend is completely different for $F_Q(\rho,D_z)$, which being enhanced for smaller values of $J_y$. This shows that the effect of entropy on metrological performance depends on the parameter being estimated. In Fig.~\ref{fig:Entropy_QFI}(b) we plot the entropy changes with the probe-state parameter $\theta$. It can be seen that the entropy is strongly affected by the choice of the initial state.  The largest entropy is obtained for $\theta=\pi/4$ whereas $\theta=0$ gives a slower increase and $\theta=\pi/2$ reaches a lower saturation value at longer times. We find that $F_Q(\rho,D_z)$ also attains its highest values around $\theta=\pi/4$, whereas $F_Q(\rho,b)$ is maximum for $\theta=0$. Therefore, the variation of the entropy with $\theta$ follows the behaviour of $F_Q(\rho,D_z)$ more closely than that of $F_Q(\rho,b)$. This suggest that the entropy can provide useful insight into the metrological properties of the system, although its relationship with the Quantum Fisher Information is not the same for different estimation tasks.

\section{Conclusion}

In this work, we have theoretically studied quantum parameter estimation in an anisotropic XYZ Heisenberg spin system with DM interaction in the presence of intrinsic decoherence. By using the QFI, we examined the estimation of both the uniform magnetic field and  DM interaction strength and explored how the exchange anisotropy, the initial probe state, and intrinsic decoherence influence the estimation precision. Our results show that there is no single set of parameters that is optimal for both estimation tasks. Instead, the best performance depends on the quantity being estimated, and a suitable choice of the exchange anisotropy and the initial state can noticeably improve the achievable precision. We also found that, although intrinsic decoherence gradually reduces the sensitivity, its impact can be alleviated to a large extent through proper parameter tuning. To obtain a broader picture of the metrological behaviour, we further compared the Quantum Fisher Information with concurrence, quantum coherence, and von Neumann entropy. These quantities help reveal different aspects of the system dynamics, but they do not always vary in the same way as the estimation precision. Overall, our study shows that anisotropic Heisenberg spin systems with DM interaction provide a useful setting for controllable quantum metrology and may serve as promising candidates for future spin-based quantum sensing applications.

\end{document}